\documentclass[preprint,amsmath,amssymb,aip,jmp]{revtex4-1}

\usepackage{times}
\usepackage{graphicx}
\usepackage{dcolumn}
\usepackage{bm}
\usepackage{soul}

\begin{document}

\title{Role of solvation in pressure-induced helix stabilization}

\author{Robert B. Best}
\email{robertbe@helix.nih.gov}
\affiliation{Laboratory of Chemical Physics,
National Institute of Diabetes and Digestive
and Kidney Diseases, National Institutes of Health, Bethesda, MD 20892-0520, U.S.A.}
\author{Cayla Miller}
\affiliation{Department of Chemical and Biomolecular Engineering, Lehigh University, Bethlehem, PA 18015, U.S.A.}
\author{Jeetain Mittal}
\email{jeetain@lehigh.edu}
\affiliation{Department of Chemical and Biomolecular Engineering, Lehigh University, Bethlehem, PA 18015, U.S.A.}

\begin{abstract}
In contrast to the well-known destabilization of
globular proteins by high pressure, recent work has shown that pressure stabilizes 
the formation of isolated $\alpha$-helices. However all simulations to
date have obtained a qualitatively opposite result within the experimental 
pressure range. We 
show that using a protein force field (Amber03w) parametrized in conjunction with
an accurate water model (TIP4P/2005) recovers
the correct pressure-dependence and an overall stability diagram for 
helix formation similar to that from experiment; on the
other hand, we confirm that using TIP3P water results in a very 
weak pressure destabilization of helices. By carefully analyzing
the contributing factors, we show that this is not merely a
consequence of different peptide conformations sampled using TIP3P. 
Rather, there is a critical role for the solvent itself in determining
the dependence of total system volume (peptide and solvent) on helix content. 
Helical peptide structures exclude a smaller volume to water,
relative to non-helical structures with both the water models, but the total system volume 
for helical conformations is higher than non-helical conformations with TIP3P 
water at low to intermediate pressures, in contrast to TIP4P/2005 water. 
Our results
further emphasize the importance of using an accurate water model
to study protein folding under conditions away from standard temperature
and pressure.
\end{abstract}

\maketitle

\section{Introduction}

The dependence of protein folding equilibria on thermodynamic
control variables can
provide important insights into the fundamental forces which stabilize
folded structures~\cite{dill1990dominant}. Variation of temperature is the most commonly 
used approach due to the ease with which this can be achieved.
A detailed comparison of the temperature-dependence of protein
stability with that of the hydrophobic effect, for example, strongly
supports the role of hydrophobic interactions in stabilizing
protein structure~\cite{baldwin-1986,robertson-1997}. 

A variable which is less exploited is pressure, due to the 
greater demands of the experiments required~\cite{dumont2009reaching,larios-2010}. 
However, 
it is well-known that high pressure tends to destabilize protein native 
structures~\cite{mozhaev1996high,silva1993pressure}, indicative of a positive change of reaction volume for
folding. A possible resolution of this initially counterintuitive 
effect \cite{kauzmann-1987} was proposed to be the existence
of cavities within the folded protein \cite{hummer-1998-3,frye-1998},
with substantial support for this hypothesis coming from experiments
on cavity-forming mutants \cite{frye-1998,roche-2012}. It was also proposed that 
high pressures can lead to water penetration of protein's hydrophobic core due to reduced 
solvent-solute interfacial free energy~\cite{cheung2006heteropolymer}.

However, the pressure-dependence of the folding equilibria for small,
independently folding elements of secondary structure such as hairpins and
helices clearly cannot fit the same picture. Since they contain no evident
internal cavities or well-defined hydrophobic core, any pressure dependence would have to 
come from other effects, which
might be obscured when studying the overall folding of a globular protein.
Recent experimental work on the pressure dependence of the helix-coil
equilibrium, using either FTIR spectroscopy\cite{takekiyo-2005,imamura-2008} or triplet state
quenching experiments\cite{neumaier-2013} has in fact found the opposite trend to that
for protein folding: namely a negative reaction volume for helix formation,
resulting in pressure stabilization of helices for all positive pressures.
The origin of this effect is not clear, especially considering the small
magnitude of the volume change ($\approx$ 0.2--1.0 cm$^3$.mol$^{-1}$ per residue). 
Molecular simulation could potentially help to
explain the origin of this result; however, simulation studies of the pressure
dependence of helix formation have qualitatively contradicted experimental
results, finding instead 
helix destabilization at low to intermediate pressures, only turning over 
to stabilization at very high
pressure \cite{paschek-2005,hatch-2014,mori-2014}. 

Here, we investigate the pressure dependence of helix formation for a model
15-residue helix-forming peptide using two different force field combinations:
the Amber ff03* protein force field\cite{best-2009-2} together with explicit
TIP3P water \cite{jorgensen-1983} and the Amber ff03w protein force field
\cite{best-2010-3} with the TIP4P/2005 water model \cite{abascal-2005}.
In agreement with earlier studies with TIP3P water on the effect of pressure 
on helix formation, we find a positive reaction volume for helix formation at 
low to intermediate pressures.
In contrast, a qualitatively correct result is obtained with TIP4P/2005, 
i.e., a negative reaction volume for helix formation. This difference, obtained
for the same sequence with almost identical protein force fields, suggests
a key role for water.  
There are essentially two ways in which water can be 
envisaged
to influence the reaction volume: (i) it may alter the conformations sampled
for a given total number of helical residues, particularly for the 
non-helical states and (ii) different water models may be more or less 
closely coordinated with helical than with non-helical structures. 
While there is no doubt that water influences
the conformational sampling, we show that in fact the different 
solvation of helical and non-helical states plays a key role in determining
the volume changes, and for both water models opposes the much larger
decrease in volume excluded to water associated with helix formation. 
The result obtained with TIP3P water at low to intermediate 
pressures is qualitatively inconsistent with experiment, as the total system 
volume is higher for helical structures than for non-helical structures.
Our results highlight the importance of using an accurate water model
for capturing biomolecular equilibria at state points away from 
standard conditions.

\section{Methods}

\subsection{Simulation methods}

Replica exchange molecular dynamics (REMD) simulations with gromacs 4.0.7 or 4.5.3
\cite{hess-2008} were used to sample the folding of the blocked
peptide Ac-(AAQAA)$_3$-NH$_2$ with either (i) the Amber ff03* force
field\cite{best-2009-2} and TIP3P water model\cite{jorgensen-1983} or 
(ii) the Amber ff03w force field \cite{best-2010-3} and
the TIP4P/2005 water model \cite{abascal-2005}. For ff03*, 42 replicas 
spanning a temperature
range from 275 to 496 K were used, and for ff03w, 40 replicas spanning 
250 to 452 K. Periodic boundary conditions with a truncated octadron
minimum image cell of initial edge 4.5 nm were used. Simulations 
were maintained at a constant pressure with a 
Parrinello-Rahman barostat\cite{parrinello-1981} and temperature was controlled via
Langevin dynamics with a friction coefficient of 0.2 ps$^{-1}$ (ff03*)
or 1.0 ps$^{-1}$ (ff03w). Note that the difference in friction coefficients
will not alter the equilibrium properties considered here, and both 
values are sufficient to maintain constant temperature. Simulations
were run for 100-200 ns per replica. A seperate set of REMD runs was 
performed for each pressure, using 1 bar, 2 kbar, 4 kbar, 8 kbar and
12 kbar for Amber ff03* and 1 bar , 1 kbar, 4 kbar, 8 kbar for Amber ff03w.

To determine the change of volume associated with helix formation, initial
configurations were obtained by random selection from the 298 K replica
of the ff03* simulations. For each number of helical residues ($0,3,4,\ldots 15$),
20 configurations were chosen with that helix content. From each of these
configurations, 10 ns simulations were
performed using position restraints to keep the peptide close to its initial 
structure, with force constants of 1000 kJ.mol$^{-1}$.nm$^{-2}$ on each
cartesian coordinate, and using
both the ff03* and ff03w force fields. Average total system volumes were
computed from each of these runs. 

\subsection{Helix formation}

We define helical states using the backbone Ramachandran angles, in the
spirit of the Lifson-Roig theory. We define the helical region of the Ramachandran
map as $\phi \in [-100^\circ,-30^\circ]$ and $\psi \in [-67^\circ,-7^\circ]$.
A helical segment is defined as three consecutive residues with their backbone
torsion angles lying in this range. Therefore the total number of helical residues
in the blocked 15-residue peptide considered can take on values in $\{0,3,4,5,\ldots,14,15\}$.
This method gives results which are consistent with other definitions of 
helix \cite{best-2009-2}. 

\subsection{Helix-coil models}

Simulation
data on helix formation were initially described by a Lifson-Roig model \cite{lifson-1961}, with
nucleation parameter $v$ and elongation parameter $w$. These parameters were
fitted by maximizing the likelihood of the observed conformations, given the
equilibrium probabilities generated from the Lifson-Roig partition function,
using a previously described procedure \cite{best-2009-2}. Since the experimental
data had been fitted to the Zimm-Bragg model for helix formation \cite{zimm-1959},
we also converted the fit parameters to the Zimm-Bragg nucleation and elongation
parameters $\sigma$ and $s$ respectively, using the approximate relations \cite{qian-1992}:
\begin{eqnarray}
\sigma & \approx & \frac{v^2}{(1+v)^4}, \nonumber \\
s & \approx & \frac{w}{1+v}.
\end{eqnarray}

\subsection{Thermodynamic Model}

The dependence of the helix elongation free energy, $\Delta G_\mathrm{el}$ on
pressure and temperature was fitted to a thermodynamic model:
\begin{eqnarray}
\Delta G_\mathrm{el}(P,T) & =&  \Delta H_0 + \Delta C_P(T-T_0) \nonumber \\
			&    & -T\Delta S_0 - T\Delta C_P\ln(T/T_0) \nonumber \\
			&    & +\Delta V(P-P_0) + \frac{\Delta\beta}{2}(P-P_0)^2 \nonumber  \\
			&    & +\Delta \alpha(T-T_0)(P-P_0).
\label{whatsit}
\end{eqnarray}
In this expression, $\Delta H_0$, $\Delta S_0$, and $\Delta V_0$ are 
the change of enthalpy, entropy and volume as a result of helix
elongation at reference conditions, taken to be $T_0=298$ K and $P_0=1$ bar pressure.
In addition, a constant change of heat capacity, $\Delta C_P$, change
of linear expansion coefficient $\Delta \alpha$ and change of compressibility,
$\Delta \beta$ were assumed. The data were fitted by non-linear least squares,
and errors were estimated by Monte Carlo bootstrapping. 

\section{Results}

In order to determine the effects controlling helix stability under pressure,
we carefully selected two protein force fields, Amber ff03* and Amber ff03w.
These models are almost identical, being originally based on the standard
Amber ff03 force field\cite{duan-2003}. They only differ in that an additional 
empirically determined Fourier term has been added to the $\psi$ backbone torsion angle in each 
case, in order to approximately match experimental helix propensities
near 300 K~\cite{best-2009-2,best-2010-3}. In the case of ff03*, the calibration was done in 
conjunction 
with the TIP3P water model, while for ff03w it was done with TIP4P/2005 water
(below, it will be assumed when discussing Amber ff03* and ff03w 
that the TIP3P and TIP4P/2005 water models were used, respectively).
This allows us to test specifically the effects of the water model using 
closely related protein force fields that have very similar helical populations 
under standard conditions of temperature and pressure. We study the 
15-residue peptide Ac-(AAQAA)$_3$-NH$_2$,
as a model system which is known to form helix at low temperature
\cite{shalongo-1994}, and which has been extensively characterized in
previous simulations \cite{best-2009-2,best-2010-3,best-2012-2}.

\begin{figure}[tbhp]
\centering
\begin{tabular}{c}
\includegraphics[width=7cm]{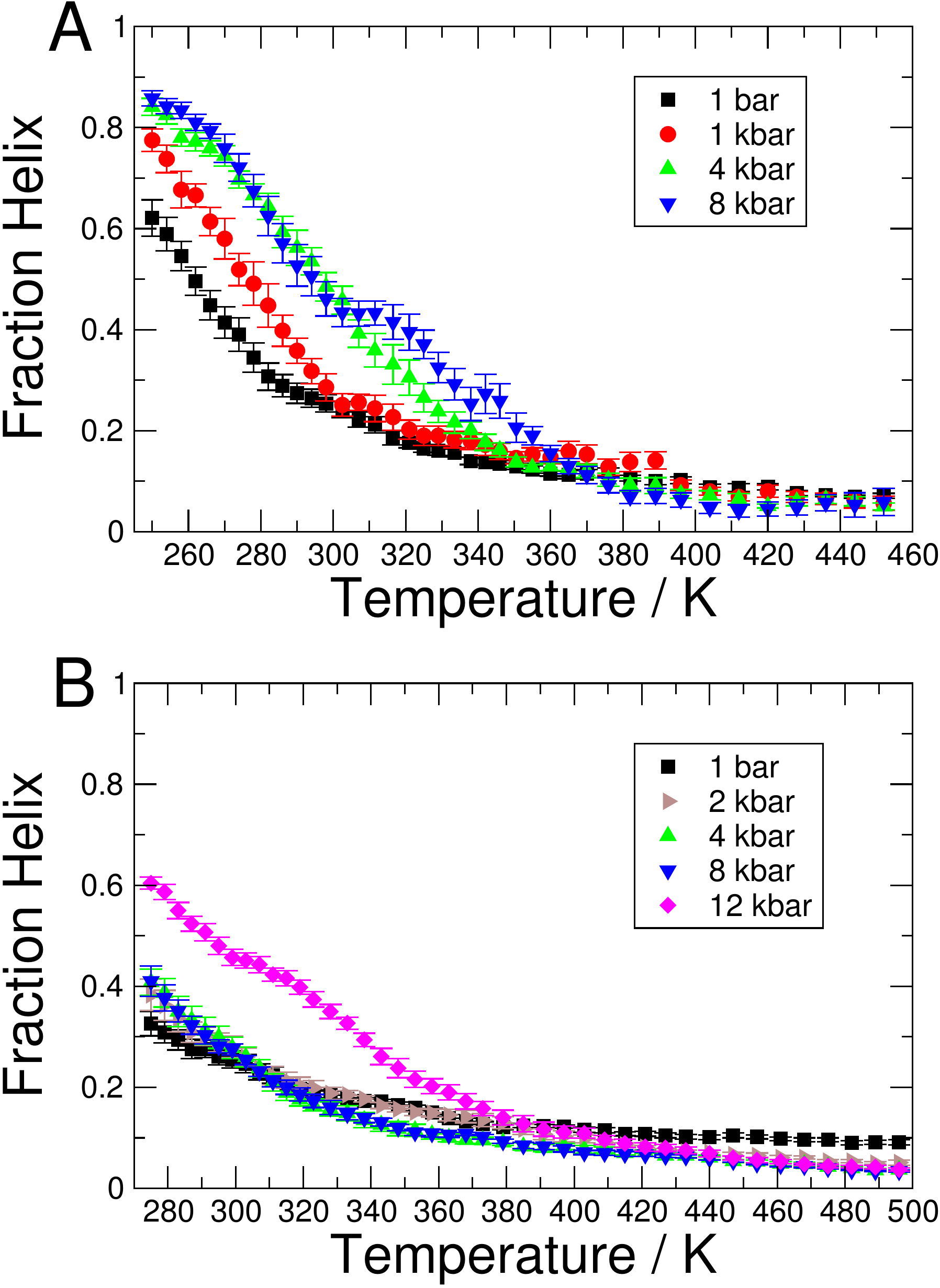}
\end{tabular}
\caption{Pressure-dependence of fraction helix.
(A) Amber ff03w, TIP4P/2005 water; (B) Amber ff03*, TIP3P water.}
\label{fhelix-fig}
\end{figure}

Replica exchange molecular dynamics (REMD) simulations were performed in order to
sample the temperature-dependent helix-coil equilibrium. The average 
fraction helix at each temperature is shown in Figure \ref{fhelix-fig},
as determined using backbone dihedral angles. Very similar results are
obtained using the standard DSSP \cite{kabsch-1983} algorithm, which instead uses backbone
hydrogen bonds  (See Fig. S1 \cite{suppinfo}).
As expected both Amber ff03* and ff03w populate $20-30$ \% helix at
300 K and 1 bar pressure, where they were parametrized against experimentally 
determined helix populations. For Amber ff03w,
all pressures
used resulted in a stabilization of the peptide, even at 1 kbar. On
the other hand, over a wide range, from 1 bar to 8 kbar,
pressure had very little effect
on the overall helix propensity for Amber ff03*: a slight decrease in helix fraction at temperatures
greater than $\sim 300 K$ is observed. Only for a pressure of 12 kbar is a significant
stabilization obtained. Qualitatively, these results suggest a negative reaction 
volume for helix formation using Amber ff03w, but a very small or 
positive reaction volume at low to intermediate pressure using Amber ff03*. 

We quantify the effect of pressure on the helix-coil equilibrium using
a thermodynamic model. However, helix-formation is not a simple two-state process,
and involves a broad spectrum of populated intermediates. Therefore, we first fit
a simple Ising-like statistical mechanics model which can capture the 
helix-coil transition. The two classic partition functions for helix-coil formation
are those by Zimm and Bragg \cite{zimm-1959} and by Lifson and Roig \cite{lifson-1961},
which are approximately equivalent \cite{qian-1992}. We have 
determined parameters for both models here. The data were initially
fitted to the Lifson-Roig model, using a previously described
maximum likelihood method \cite{best-2009-2}, yielding a nucleation 
parameter $v$ and an elongation parameter $w$. 
These parameters were then converted into the corresponding
parameters for the Zimm-Bragg model, $\sigma$ and $s$. Overall, all
these parameters show essentially similar trends to
the global fraction of helix, but can be more justifiably fitted
to a thermodynamic two-state model since they describe the
microscopic transitions involving the flipping of individual 
residues between helical and extended conformations. Below we focus 
on the Zimm-Bragg model,
as this has been used to characterize the experimental data
\cite{imamura-2008,neumaier-2013}. In particular, the elongation
parameter $s$ corresponds to the equilibrium constant for adding
a single helical hydrogen bond. 

\begin{figure}[tbhp]
\centering
\begin{tabular}{c}
\includegraphics[width=7cm]{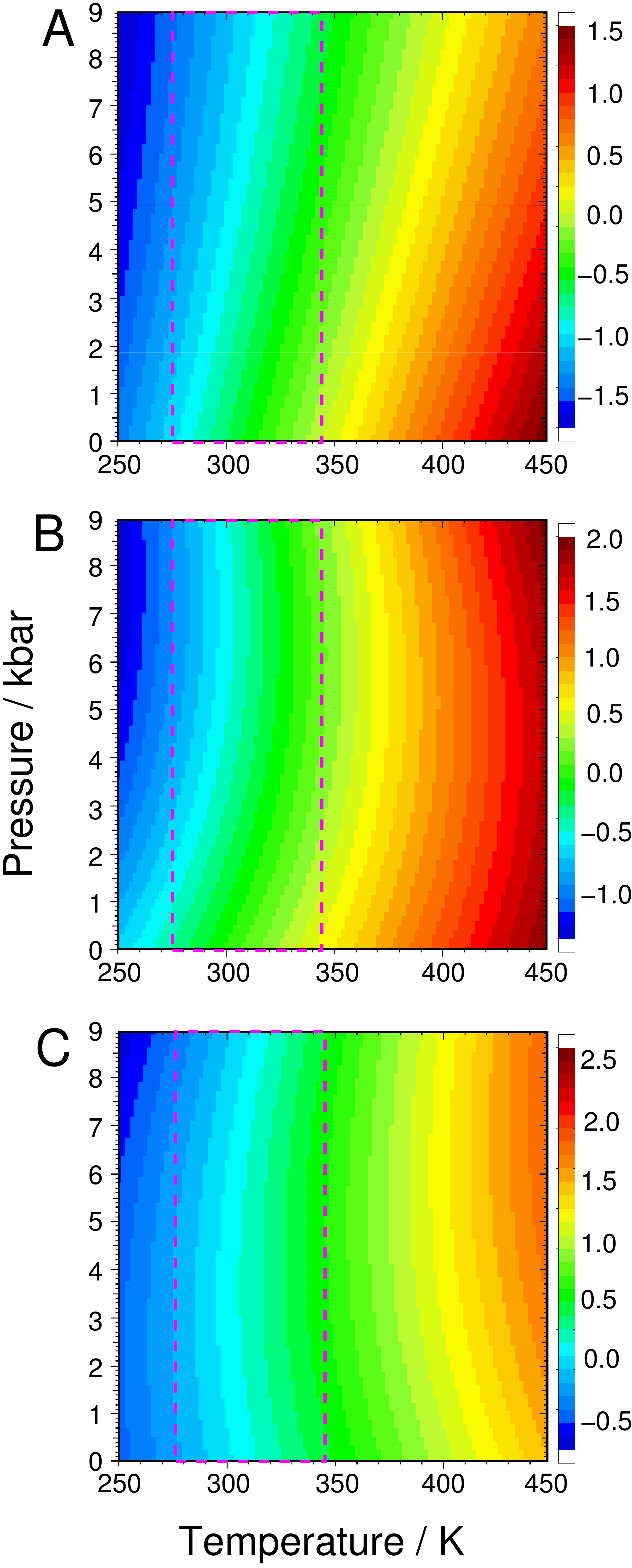}
\end{tabular}
\caption{Stability diagrams for helix elongation. The free energy
associated with elongating a helix by one residue, $\Delta G_\mathrm{el}$
is obtained from the Zimm-Bragg\cite{zimm-1959} elongation parameter $s$ as 
$\Delta G_\mathrm{el} = -RT\ln s$, as a function of temperature
and pressure. Shown are the stability diagrams obtained by fitting
Eq. \ref{whatsit} to (A) experimental $RT\ln s(P,T)$ obtained from 
Imamura and Katu \cite{imamura-2008} for the peptide 
Ac-AA(AAKAA)$_3$AAY-NH$_2$; (B) constant-pressure replica
exchange simulations of the peptide Ac-(AAQAA)$_3$-NH$_2$ with the 
ff03w force field\cite{best-2010-3} and TIP4P/2005 water
model \cite{abascal-2005}; (B) constant-pressure replica
exchange simulations of Ac-(AAQAA)$_3$-NH$_2$ with the 
ff03* force field\cite{best-2009-2} and TIP3P water
model \cite{jorgensen-1983}. Dashed magenta box indicates
approximate region covered by experimental data.
Energy units are kJ.mol$^{-1}$.}
\label{stab-diag}
\end{figure}

We fitted a thermodynamic model to the elongation free energy, here
defined as $\Delta G_\mathrm{el}(P,T) = -RT\ln s(P,T)$. The model
includes changes of enthalpy $\Delta H_0$, entropy $\Delta S_0$
and reaction volume $\Delta V_0$ under standard conditions, as
well as constant differences in heat capacity $\Delta C_P$, isothermal 
compressibility $\Delta\beta$ and linear thermal expansion coefficient 
$\Delta\alpha$ to describe the temperature- and pressure-dependence. The
same model was fitted to the data for $RT\ln s(P,T)$ reported by
Imamura \textit{et al.}, based on FTIR measurements \cite{imamura-2008}.
The fits to the raw data are shown in Fig. S2 \cite{suppinfo}. In Fig.
\ref{stab-diag} we show the stability diagram for each
force field and for experiment; the fitted parameters are listed
in Table \ref{fitparms}. 

Overall, the stability diagram obtained with the Amber ff03w 
force field is very similar to experiment, bearing in mind that
the experiments were done on a different alanine-based peptide
(AK20) -- we note that the stability diagrams for the AK16 peptide obtained
by Hatch \textit{et al}\cite{hatch-2014} with Amber ff03* and 
TIP3P are qualitatively very similar to those we obtain for 
Ac-(AAQAA)$_3$-NH$_2$ with the same force field and water model.
This is particularly true in the range of temperature
and pressure probed by the experiments, indicated by the broken
lines in Fig. \ref{stab-diag}. In both cases, increasing pressure
clearly stabilizes helical states, as also reflected in the negative
reaction volume of $-0.8$ cm$^3$.mol$^{-1}$ for experiment and
$-1.5$ cm$^3$.mol$^{-1}$ for Amber ff03w (Table 1). It also captures 
quite well the overall enthalpy and entropy changes for adding
a helical residue, as previously noted \cite{best-2010-3}. In 
contrast, the stability diagram for the ff03* force field 
differs in some important respects. Application of low pressures
will have little effect on, or slightly destabilize helical,
reflected in the small positive $\Delta V_0 = 0.4$ cm$^3$.mol$^{-1}$ (Table 1)
for helix formation.
Additionally, the changes in enthalpy and entropy are almost 
half the experimental estimates, indicative of too low a cooperativity
of helix melting\cite{best-2009-2}.  For experiment and both force
fields, the changes in heat capacity are small. This is in contrast to 
protein folding \cite{robertson-1997}, and may indicate a limited role for the hydrophobic
effect in stabilizing helices. For both force fields, the difference
in thermal linear expansion coefficients, $\Delta\alpha$, is small
and positive, $\sim 4$ $\mathrm{cm}^3.\mathrm{mol}^{-1}.\mathrm{K}^{-1}$.
This implies a change of reaction volume with temperature which will 
slightly increase the positive reaction volume for Amber ff03*, but
is insufficient to change the sign of the negative reaction volume
using Amber ff03w. Over the the range of temperature where liquid 
water is stable
the effect is too small to qualitatively
change the pressure-dependence of the two models.

\begin{figure}[tbhp]
\centering
\begin{tabular}{c}
\includegraphics[width=7cm]{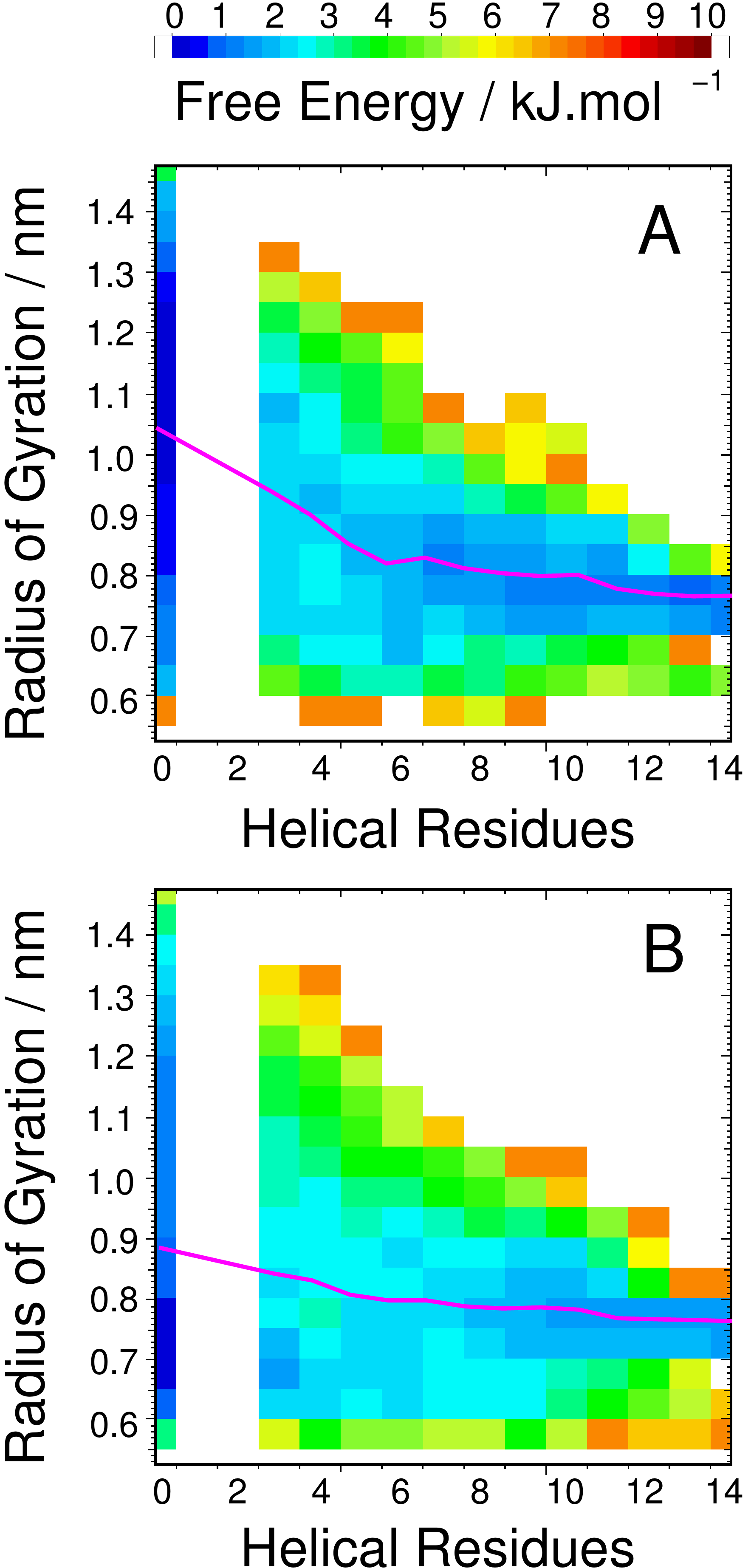}
\end{tabular}
\caption{Free energy surfaces for radius of gyration and number
of helical residues. (A) Amber ff03w, (B) Amber ff03*. Magenta
line indicates mean radius of gyration for a given number of
helical residues.}
\label{rg-helix-fig}
\end{figure}

The differences in enthalpy
and entropy between the two solvent models have been explained
in terms of the strength of solvent interactions with the peptide
chain \cite{best-2010-3}. These differences also result in a more
expanded unfolded state for ff03w relative to ff03*. How can the differences in reaction 
volume be understood? The first explanation would be in terms
of changes in the peptide free energy surface -- i.e. different 
conformations are preferred by each force field, particularly 
for non-helical states. This is undoubtedly 
a contribution, since these free energy surfaces are evidently
different. In Fig. \ref{rg-helix-fig}, we show representative
two-dimensional free energy surfaces as a function of the radius
of gyration and the number of helical residues. 
As anticipated, the radius of gyration of the helical
states is similar, but the non-helical states are much more collapsed 
in the case of ff03* compared to ff03w.  If one supposes that a 
more collapsed unfolded state is associated with a smaller volume,
then this would be consistent 
with the observed differences in
reaction volume for helix formation for the two force fields.

\begin{figure}[tbhp]
\centering
\begin{tabular}{c}
\includegraphics[width=7cm]{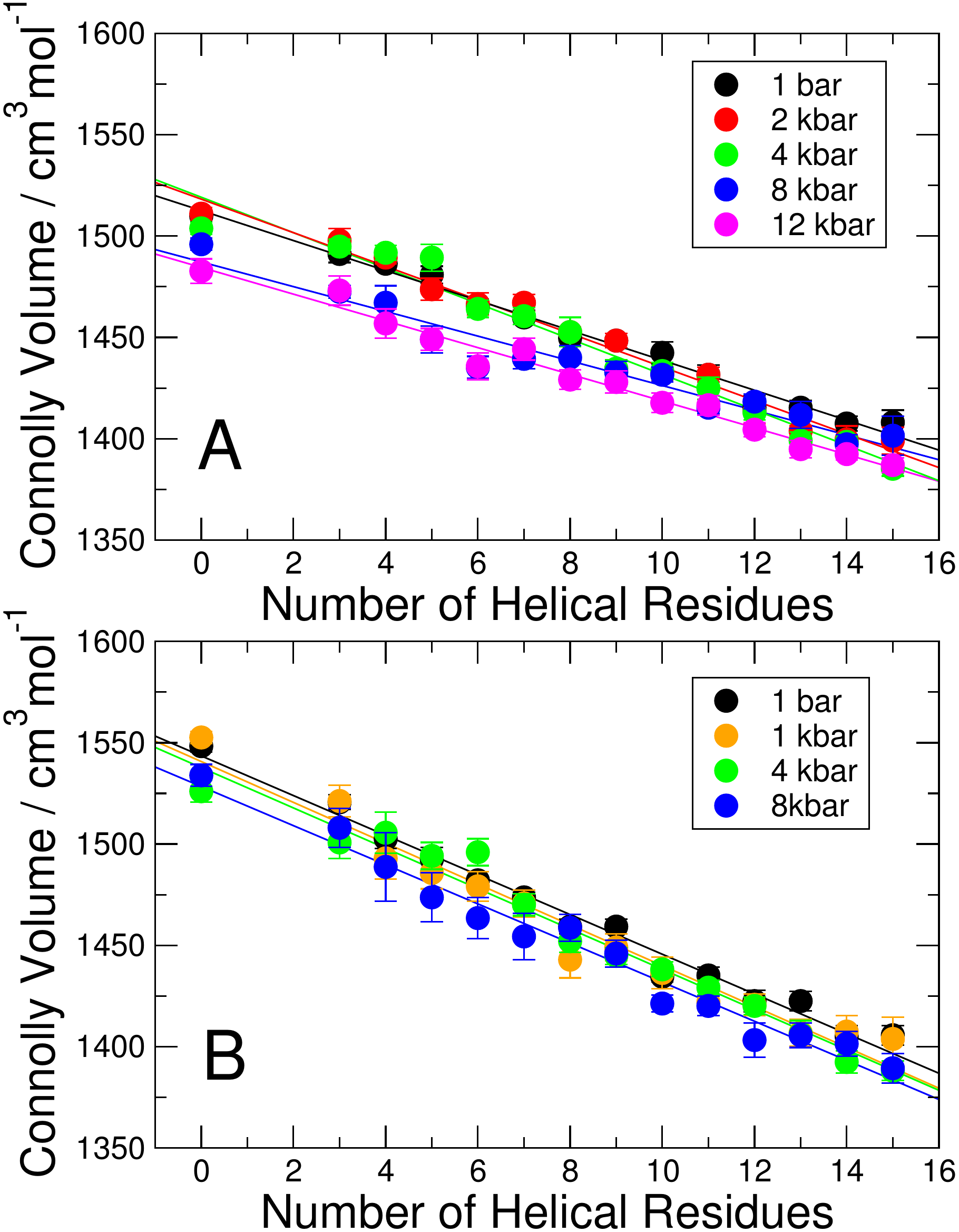}
\end{tabular}
\caption{Dependence of Connolly volume on helicity. The 
Connolly volume averaged over configurations with the same number of helical
residues is shown for (A) Amber ff03* and (B) Amber ff03w.}
\label{convol-fig}
\end{figure}

In order to investigate the above hypothesis, we determined the approximate
volume excluded to water by each peptide conformation by using the Connolly
volume: this is the volume which is inaccessible to a sphere of radius 0.14 nm.
This is the simplest way in which the volume of a configuration can be
estimated. The average Connolly volumes for configurations with the 
same number of helical residues are shown in Fig. \ref{convol-fig},
for the replica at 298 K. The
results clearly show that more helical states have a smaller total 
volume for both ff03* and ff03w, which is qualitatively consistent with the negative 
change of volume for
helix formation observed experimentally. However, it does not explain
why under low pressure conditions the reaction volume for TIP3P may
be positive, and the overall changes of 10-15 cm$^3$.mol$^{-1}$ per
helical residue are about an order of magnitude larger than the
volume changes estimated from the thermodynamic fits in Table 
\ref{fitparms}. It also does not explain the reduction of reaction
volume for ff03* with pressure, such that it becomes negative at 
sufficiently high pressures. Although 
we find
that helical states do have smaller Connolly volume above 4 kbar,
in agreement with earlier findings based on the radius of gyration
\cite{mori-2014}, we also find an even greater reduction
in volume for non-helical states so that the difference
in volume between non-helical and helical states
would become more positive with increasing pressure, the 
opposite trend to that observed
for the total reaction volume of the system. Therefore, a simple picture based only
on properties of the peptide configurations does not tell the
whole story.

\begin{figure}[tbhp]
\centering
\begin{tabular}{c}
\includegraphics[width=7cm]{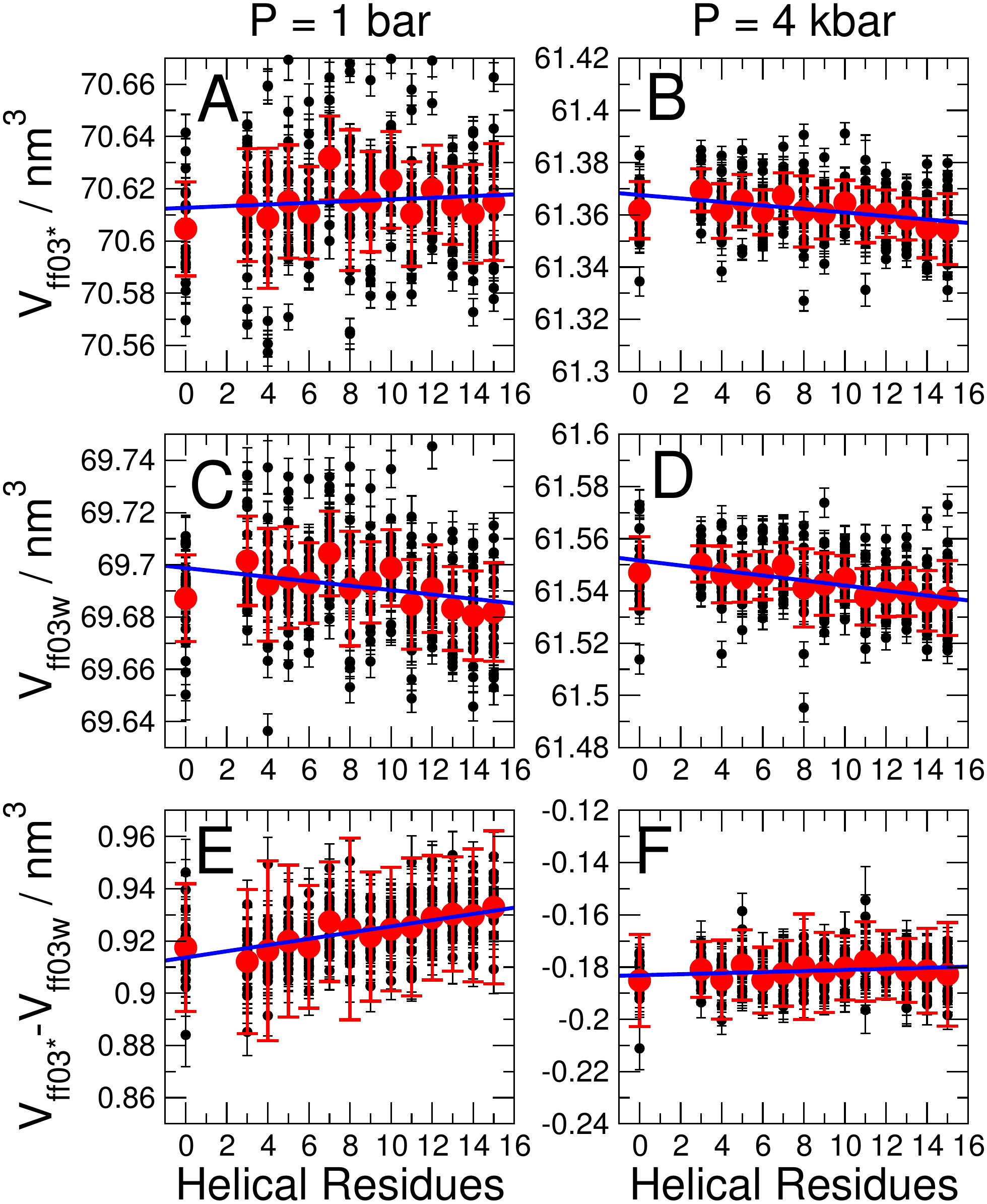}
\end{tabular}
\caption{System volumes determined for a common set of peptide configurations.
(A) Amber ff03*, 1 bar; (B) Amber ff03*, 4 kbar; (C) Amber ff03w, 1 bar; (D) Amber ff03w, 4 kbar; 
(E), (F) difference between volumes using Amber ff03w and Amber ff03* at 1 bar and 4 kbar
respectively. Black data points are the average volumes determined for individual configurations
(20 per number of helical residues) and red symbols are average system volumes for all
configurations with the same number of helical residues.}
\label{restr-vol-fig}
\end{figure}

Naturally, the surrounding solvent may also play a role in determining
the dependence of system volume on peptide conformation, and
previous studies have suggested that peptide solvation changes
with increasing pressure \cite{paschek-2005,paschek-2008}, and
changes in water structure with increasing pressure are known
to alter the hydrophobic effect \cite{hummer-1998-3}.
We probed for this by randomly drawing configurations from the 
298 K replica from the 1 bar Amber ff03* REMD simulation, and then 
determining for each of
these the average system volume using different force fields
and system pressures (1 bar and 4 bar). That is, we effectively remove the 
influence of different free energy surfaces with different 
force fields by always
considering the same set of peptide configurations. By running
sufficiently long constant pressure simulations with each of these configurations,
restrained to their initial position, we can accurately determine
the average system volume, as a function of the number of helical
residues. The results of these simulations for the Amber ff03* and
ff03w force fields are summarized
in Fig. \ref{restr-vol-fig}. We find that this simple
analysis captures the reaction volume effects inferred from the 
thermodynamic fits. Namely, at 1 bar pressure, the reaction 
volume for helix formation is positive with ff03* and negative
with ff03w (Fig. \ref{restr-vol-fig}A,C), while at 4 bar pressure
the change of volume upon helix formation becomes negative even
for TIP3P (Fig. \ref{restr-vol-fig}B,D). The large scatter in 
the individual system volumes (black data points) indicates that
other factors besides helicity are important for determining system
volume. Nonetheless, when the volume of a given configuration 
with ff03w is subtracted from that with ff03*, these effects
are largely eliminated, leaving helicity as the main determinant
of the difference between the two water models (Figure \ref{restr-vol-fig}E-F).
This result highlights the importance of solvation in 
determining the difference in apparent volume between the helical 
and less helical states. 

Although Amber ff03* with the TIP3P water model may fail to describe the correct
qualitative pressure-dependence of helix stability, it should
be noted that the differences in reaction volume under consideration
are extremely tiny: a change of volume 
$\Delta V = 1 ~~ \mathrm{cm}^3.\mathrm{mol}^{-1}$ per
residue, or $1.66 \times 10^{-3}$ nm$^3$ per molecule per residue can be 
related to a change of apparent helix radius $\Delta R$ using
a helix pitch of $\Delta L \sim 0.15$ nm and helix radius 
$R \sim 0.45$ nm via $\Delta R \approx \Delta V / 2\pi R\Delta L$,
yielding a change of apparent radius $\Delta R \approx 0.0039$ nm.
While this is a simplified model calculation, it serves to illustrate
the subtlety of the effect that must be captured.

\section{Conclusion}

The dependence of globular protein stability on pressure has been found to 
be mainly determined by cavities in the folded structure, masking other 
possible effects of interest. In the case of helix formation, 
in fact the opposite trend is found to that for protein folding,
namely pressure stabilization of helices.
We have shown here that simulations with an accurate water model,
TIP4P/2005, 
are capable of capturing the pressure dependence of helix formation.
Further, in agreement with earlier work, we find that using the TIP3P 
water model leads to pressure-induced destabilization of helices, 
a qualitatively incorrect result. We further show here that this difference is not
merely due to the different peptide conformations sampled with that
water model. Instead there is a critical role for solvent structure
in determining the reaction volume changes. 
Taken together, our results emphasize the importance of using an accurate
water model for capturing the folding of peptides and proteins under
different thermodynamic conditions. 

The experimental 
results obtained for helix formation have proved to be a stringent
test for simulation models. In future, it would be very interesting
to compare simulation results with experimental data for other model
peptides such as $\beta$-hairpins, should such data become available.
Experimental kinetics results for the pressure-dependence of helix
\cite{neumaier-2013} and protein folding \cite{roche-2013,liu-2014} kinetics are 
also a rich source of information for future detailed comparison with 
molecular simulation.

\begin{acknowledgments}
RB is supported by the Intramural Research Program of the National Insitute of
Diabetes and Digestive and Kidney Diseases of the National Institutes of
Health. JM is supported by Alfred P. Sloan Foundation Research Fellowship. This study utilized the high-performance computational
capabilities of the Biowulf Linux cluster at the National Institutes of Health,
Bethesda, Md. (http://biowulf.nih.gov) and the high-performance computing capabilities of the Extreme Science and Engineering Discovery Environment (XSEDE), which is supported by the National Science Foundation grant no. TG-MCB-120014.
\end{acknowledgments}

\newpage
\section{Tables}

\begin{table}[h]
\caption{Parameters for fits of helix elongation free energy
$\Delta G_\mathrm{el}$ to thermodynamic
model (Eq. \ref{whatsit}). Numbers in brackets are the error in the
last significant figure estimated by bootstrap Monte Carlo.}
\label{fitparms}
\centering
\begin{tabular}{ l @{~~} l @{~~} r  @{~~~~} r  @{~~~~} r }
\hline
\hline
Parameter 	& Units 	& Experiment 	& ff03w & ff03* \\
\hline
$\Delta H_0$	& kJ.mol$^{-1}$	& $-5.2(3)$	& $-4.1(1)$	& $-2.9(1)$	\\
$\Delta S_0$	& J.mol$^{-1}$.K$^{-1}$	
				& $-15(1)$	& $-13.4(2)$	& $-9.7(3)$	\\
$\Delta C_P$	& J.mol$^{-1}$.K$^{-1}$	
				& $0(1)$	 & $2(1)$	& $-1(1)$\\
$\Delta V_0$	& cm$^3$.mol$^{-1}$	
				& $-0.8(1)$	& $-1.5(1)$	& $0.4(2)$	\\
$\Delta \beta$	& $\times 10^{-5}$ cm$^3$.mol$^{-1}$.bar$^{-1}$	
				& $6(3)$	&  $22(2)$	& $-13(2)$	\\
$\Delta \alpha$	& $\times 10^{-3}$ cm$^3$.mol$^{-1}$.K$^{-1}$	
				& $-3(2)$	& $4.1(6)$	& $3.4(4)$	\\
\hline
\hline
\end{tabular}
\end{table}

\newpage
\section{Figures}

\begin{thebibliography}{10}%
\makeatletter
\providecommand \@ifxundefined [1]{%
 \ifx #1\undefined \expandafter \@firstoftwo
 \else \expandafter \@secondoftwo
\fi
}%
\providecommand \@ifnum [1]{%
 \ifnum #1\expandafter \@firstoftwo
 \else \expandafter \@secondoftwo
\fi
}%
\providecommand \enquote [1]{``#1''}%
\providecommand \bibnamefont  [1]{#1}%
\providecommand \bibfnamefont [1]{#1}%
\providecommand \citenamefont [1]{#1}%
\providecommand\href[0]{\@sanitize\@href}%
\providecommand\@href[1]{\endgroup\@@startlink{#1}\endgroup\@@href}%
\providecommand\@@href[1]{#1\@@endlink}%
\providecommand \@sanitize [0]{\begingroup\catcode`\&12\catcode`\#12\relax}%
\@ifxundefined \pdfoutput {\@firstoftwo}{%
 \@ifnum{\z@=\pdfoutput}{\@firstoftwo}{\@secondoftwo}%
}{%
 \providecommand\@@startlink[1]{\leavevmode\special{html:<a href="#1">}}%
 \providecommand\@@endlink[0]{\special{html:</a>}}%
}{%
 \providecommand\@@startlink[1]{%
  \leavevmode
  \pdfstartlink
   attr{/Border[0 0 1 ]/H/I/C[0 1 1]}%
   user{/Subtype/Link/A<</Type/Action/S/URI/URI(#1)>>}%
  \relax
 }%
 \providecommand\@@endlink[0]{\pdfendlink}%
}%
\providecommand \url  [0]{\begingroup\@sanitize \@url }%
\providecommand \@url [1]{\endgroup\@href {#1}{\urlprefix}}%
\providecommand \urlprefix [0]{URL }%
\providecommand \Eprint[0]{\href }%
\@ifxundefined \urlstyle {%
  \providecommand \doi [1]{doi:\discretionary{}{}{}#1}%
}{%
  \providecommand \doi [0]{doi:\discretionary{}{}{}\begingroup
  \urlstyle{rm}\Url }%
}%
\providecommand \doibase [0]{http://dx.doi.org/}%
\providecommand \Doi[1]{\href{\doibase#1}}%
\providecommand \selectlanguage [0]{\@gobble}%
\providecommand \bibinfo [0]{\@secondoftwo}%
\providecommand \bibfield [0]{\@secondoftwo}%
\providecommand \translation [1]{[#1]}%
\providecommand \BibitemOpen[0]{}%
\providecommand \bibitemStop [0]{}%
\providecommand \bibitemNoStop [0]{.\EOS\space}%
\providecommand \EOS [0]{\spacefactor3000\relax}%
\providecommand \BibitemShut [1]{\csname bibitem#1\endcsname}%
\bibitem{dill1990dominant}%
  \BibitemOpen
  \bibfield{author}{%
  \bibinfo {author} {\bibfnamefont{Ken~A}\ \bibnamefont{Dill}},\ }%
  \bibfield{title}{%
  \enquote{\bibinfo {title} {Dominant forces in protein folding},}\ }%
  \bibfield{journal}{%
  \bibinfo {journal} {Biochemistry}\ }%
  \textbf{\bibinfo {volume} {29}},\ \bibinfo {pages} {7133--7155} (\bibinfo
  {year} {1990})\BibitemShut{NoStop}%
\bibitem{baldwin-1986}%
  \BibitemOpen
  \bibfield{author}{%
  \bibinfo {author} {\bibfnamefont{Robert~L.}\ \bibnamefont{Baldwin}},\ }%
  \bibfield{title}{%
  \enquote{\bibinfo {title} {Temperature dependence of the hydrophobic
  interaction in protein folding},}\ }%
  \bibfield{journal}{%
  \bibinfo {journal} {Proc. Natl. Acad. Sci. U. S. A.}\ }%
  \textbf{\bibinfo {volume} {83}},\ \bibinfo {pages} {8069--8072} (\bibinfo
  {year} {1986})\BibitemShut{NoStop}%
\bibitem{robertson-1997}%
  \BibitemOpen
  \bibfield{author}{%
  \bibinfo {author} {\bibfnamefont{Andrew~D.}\ \bibnamefont{Robertson}}\ and\
  \bibinfo {author} {\bibfnamefont{Kenneth~P.}\ \bibnamefont{Murphy}},\ }%
  \bibfield{title}{%
  \enquote{\bibinfo {title} {Protein structure and the energetics of protein
  stability},}\ }%
  \bibfield{journal}{%
  \bibinfo {journal} {Chem. Rev.}\ }%
  \textbf{\bibinfo {volume} {97}},\ \bibinfo {pages} {1251--1267} (\bibinfo
  {year} {1997})\BibitemShut{NoStop}%
\bibitem{dumont2009reaching}%
  \BibitemOpen
  \bibfield{author}{%
  \bibinfo {author} {\bibfnamefont{Charles}\ \bibnamefont{Dumont}}, \bibinfo
  {author} {\bibfnamefont{Tryggvi}\ \bibnamefont{Emilsson}},\ and\ \bibinfo
  {author} {\bibfnamefont{Martin}\ \bibnamefont{Gruebele}},\ }%
  \bibfield{title}{%
  \enquote{\bibinfo {title} {Reaching the protein folding speed limit with
  large, sub-microsecond pressure jumps},}\ }%
  \bibfield{journal}{%
  \bibinfo {journal} {Nature Methods}\ }%
  \textbf{\bibinfo {volume} {6}},\ \bibinfo {pages} {515--519} (\bibinfo {year}
  {2009})\BibitemShut{NoStop}%
\bibitem{larios-2010}%
  \BibitemOpen
  \bibfield{author}{%
  \bibinfo {author} {\bibfnamefont{Edgar}\ \bibnamefont{Larios}}\ and\ \bibinfo
  {author} {\bibfnamefont{Martin}\ \bibnamefont{Gruebele}},\ }%
  \bibfield{title}{%
  \enquote{\bibinfo {title} {Protein stability at negative pressure},}\ }%
  \bibfield{journal}{%
  \bibinfo {journal} {Methods}\ }%
  \textbf{\bibinfo {volume} {52}},\ \bibinfo {pages} {51--56} (\bibinfo {year}
  {2100})\BibitemShut{NoStop}%
\bibitem{mozhaev1996high}%
  \BibitemOpen
  \bibfield{author}{%
  \bibinfo {author} {\bibfnamefont{Vadim~V}\ \bibnamefont{Mozhaev}}, \bibinfo
  {author} {\bibfnamefont{Karel}\ \bibnamefont{Heremans}}, \bibinfo {author}
  {\bibfnamefont{Johannes}\ \bibnamefont{Frank}}, \bibinfo {author}
  {\bibfnamefont{Patrick}\ \bibnamefont{Masson}},\ and\ \bibinfo {author}
  {\bibfnamefont{Claude}\ \bibnamefont{Balny}},\ }%
  \bibfield{title}{%
  \enquote{\bibinfo {title} {High pressure effects on protein structure and
  function},}\ }%
  \bibfield{journal}{%
  \bibinfo {journal} {Proteins-Structure Function and Genetics}\ }%
  \textbf{\bibinfo {volume} {24}},\ \bibinfo {pages} {81--91} (\bibinfo {year}
  {1996})\BibitemShut{NoStop}%
\bibitem{silva1993pressure}%
  \BibitemOpen
  \bibfield{author}{%
  \bibinfo {author} {\bibfnamefont{JL}~\bibnamefont{Silva}}\ and\ \bibinfo
  {author} {\bibfnamefont{G}~\bibnamefont{Weber}},\ }%
  \bibfield{title}{%
  \enquote{\bibinfo {title} {Pressure stability of proteins},}\ }%
  \bibfield{journal}{%
  \bibinfo {journal} {Annual Review of Physical Chemistry}\ }%
  \textbf{\bibinfo {volume} {44}},\ \bibinfo {pages} {89--113} (\bibinfo {year}
  {1993})\BibitemShut{NoStop}%
\bibitem{kauzmann-1987}%
  \BibitemOpen
  \bibfield{author}{%
  \bibinfo {author} {\bibfnamefont{Walter}\ \bibnamefont{Kauzmann}},\ }%
  \bibfield{title}{%
  \enquote{\bibinfo {title} {Thermodynamics of unfolding},}\ }%
  \bibfield{journal}{%
  \bibinfo {journal} {Nature}\ }%
  \textbf{\bibinfo {volume} {325}},\ \bibinfo {pages} {763--764} (\bibinfo
  {year} {1987})\BibitemShut{NoStop}%
\bibitem{hummer-1998-3}%
  \BibitemOpen
  \bibfield{author}{%
  \bibinfo {author} {\bibfnamefont{Gerhard}\ \bibnamefont{Hummer}}, \bibinfo
  {author} {\bibfnamefont{Shekhar}\ \bibnamefont{Garde}}, \bibinfo {author}
  {\bibfnamefont{Angel~E.}\ \bibnamefont{{Garc\'ia}}}, \bibinfo {author}
  {\bibfnamefont{Michael~E.}\ \bibnamefont{Paulaitis}},\ and\ \bibinfo {author}
  {\bibfnamefont{Lawrence~R.}\ \bibnamefont{Pratt}},\ }%
  \bibfield{title}{%
  \enquote{\bibinfo {title} {The pressure dependence of hydrophobic
  interactions is consistent with the observed pressure denaturation of
  proteins},}\ }%
  \bibfield{journal}{%
  \bibinfo {journal} {Proc. Natl. Acad. Sci. U. S. A.}\ }%
  \textbf{\bibinfo {volume} {95}},\ \bibinfo {pages} {1552--1555} (\bibinfo
  {year} {1998})\BibitemShut{NoStop}%
\bibitem{frye-1998}%
  \BibitemOpen
  \bibfield{author}{%
  \bibinfo {author} {\bibfnamefont{Kelly~J.}\ \bibnamefont{Frye}}\ and\
  \bibinfo {author} {\bibfnamefont{Catherine~A.}\ \bibnamefont{Royer}},\ }%
  \bibfield{title}{%
  \enquote{\bibinfo {title} {Probing the contribution of internal cavities to
  the volume change of protein unfolding under pressure},}\ }%
  \bibfield{journal}{%
  \bibinfo {journal} {Protein Sci.}\ }%
  \textbf{\bibinfo {volume} {7}},\ \bibinfo {pages} {2217--2222} (\bibinfo
  {year} {1998})\BibitemShut{NoStop}%
\bibitem{roche-2012}%
  \BibitemOpen
  \bibfield{author}{%
  \bibinfo {author} {\bibfnamefont{Julien}\ \bibnamefont{Roche}}, \bibinfo
  {author} {\bibfnamefont{Jose~A.}\ \bibnamefont{Caro}}, \bibinfo {author}
  {\bibfnamefont{Douglas~R.}\ \bibnamefont{Norberto}}, \bibinfo {author}
  {\bibfnamefont{Philippe}\ \bibnamefont{Barthe}}, \bibinfo {author}
  {\bibfnamefont{Christian}\ \bibnamefont{Roumestand}}, \bibinfo {author}
  {\bibfnamefont{Jamie~L.}\ \bibnamefont{Schlessman}}, \bibinfo {author}
  {\bibfnamefont{Angel~E.}\ \bibnamefont{Garcia}}, \bibinfo {author}
  {\bibfnamefont{Bertrand~E.}\ \bibnamefont{Garcia-Moreno}},\ and\ \bibinfo
  {author} {\bibfnamefont{Catherine~A.}\ \bibnamefont{Royer}},\ }%
  \bibfield{title}{%
  \enquote{\bibinfo {title} {Cavities determine the pressure unfolding of
  proteins},}\ }%
  \bibfield{journal}{%
  \bibinfo {journal} {Proc. Natl. Acad. Sci. U. S. A.}\ }%
  \textbf{\bibinfo {volume} {109}},\ \bibinfo {pages} {6945--6950} (\bibinfo
  {year} {2012})\BibitemShut{NoStop}%
\bibitem{cheung2006heteropolymer}%
  \BibitemOpen
  \bibfield{author}{%
  \bibinfo {author} {\bibfnamefont{Jason~K}\ \bibnamefont{Cheung}}, \bibinfo
  {author} {\bibfnamefont{Pooja}\ \bibnamefont{Shah}},\ and\ \bibinfo {author}
  {\bibfnamefont{Thomas~M}\ \bibnamefont{Truskett}},\ }%
  \bibfield{title}{%
  \enquote{\bibinfo {title} {Heteropolymer collapse theory for protein folding
  in the pressure-temperature plane},}\ }%
  \bibfield{journal}{%
  \bibinfo {journal} {Biophysical journal}\ }%
  \textbf{\bibinfo {volume} {91}},\ \bibinfo {pages} {2427--2435} (\bibinfo
  {year} {2006})\BibitemShut{NoStop}%
\bibitem{takekiyo-2005}%
  \BibitemOpen
  \bibfield{author}{%
  \bibinfo {author} {\bibfnamefont{Takahiro}\ \bibnamefont{Takekiyo}}, \bibinfo
  {author} {\bibfnamefont{Akio}\ \bibnamefont{Shimizu}}, \bibinfo {author}
  {\bibfnamefont{Minoru}\ \bibnamefont{Kato}},\ and\ \bibinfo {author}
  {\bibfnamefont{Yoshihiro}\ \bibnamefont{Taniguchi}},\ }%
  \bibfield{title}{%
  \enquote{\bibinfo {title} {{Pressure tuning FT-IR spectroscopic study on the
  helix-coil transition of Ala-rich oligopeptide in aqueous solution}},}\ }%
  \bibfield{journal}{%
  \bibinfo {journal} {Biochim. Biophys. Acta}\ }%
  \textbf{\bibinfo {volume} {1750}},\ \bibinfo {pages} {1--4} (\bibinfo {year}
  {2005})\BibitemShut{NoStop}%
\bibitem{imamura-2008}%
  \BibitemOpen
  \bibfield{author}{%
  \bibinfo {author} {\bibfnamefont{Hiroshi}\ \bibnamefont{Imamura}}\ and\
  \bibinfo {author} {\bibfnamefont{Minoru}\ \bibnamefont{Kato}},\ }%
  \bibfield{title}{%
  \enquote{\bibinfo {title} {{Effect of pressure on helix-coil transition of an
  alanine-based peptide: An FTIR study}},}\ }%
  \bibfield{journal}{%
  \bibinfo {journal} {Proteins}\ }%
  \textbf{\bibinfo {volume} {75}},\ \bibinfo {pages} {911--918} (\bibinfo
  {year} {2008})\BibitemShut{NoStop}%
\bibitem{neumaier-2013}%
  \BibitemOpen
  \bibfield{author}{%
  \bibinfo {author} {\bibfnamefont{Sabine}\ \bibnamefont{Neumaier}}, \bibinfo
  {author} {\bibfnamefont{Maren}\ \bibnamefont{{B\"uttner}}}, \bibinfo {author}
  {\bibfnamefont{Annett}\ \bibnamefont{Bachmann}},\ and\ \bibinfo {author}
  {\bibfnamefont{Thomas}\ \bibnamefont{Kiefhaber}},\ }%
  \bibfield{title}{%
  \enquote{\bibinfo {title} {Transition state and ground state properties of
  the helix–coil transition in peptides deduced from high-pressure
  studies},}\ }%
  \bibfield{journal}{%
  \bibinfo {journal} {Proc. Natl. Acad. Sci. U. S. A.}\ }%
  \textbf{\bibinfo {volume} {110}},\ \bibinfo {pages} {20988--20993} (\bibinfo
  {year} {2013})\BibitemShut{NoStop}%
\bibitem{paschek-2005}%
  \BibitemOpen
  \bibfield{author}{%
  \bibinfo {author} {\bibfnamefont{Dietmar}\ \bibnamefont{Paschek}}, \bibinfo
  {author} {\bibfnamefont{S.}~\bibnamefont{Gnanakaran}},\ and\ \bibinfo
  {author} {\bibfnamefont{Angel~E.}\ \bibnamefont{Garcia}},\ }%
  \bibfield{title}{%
  \enquote{\bibinfo {title} {Simulations of the pressure and temperature
  unfolding of an {$\alpha$}-helical peptide},}\ }%
  \bibfield{journal}{%
  \bibinfo {journal} {Proc. Natl. Acad. Sci. U. S. A.}\ }%
  \textbf{\bibinfo {volume} {102}},\ \bibinfo {pages} {6765--6770} (\bibinfo
  {year} {2005})\BibitemShut{NoStop}%
\bibitem{hatch-2014}%
  \BibitemOpen
  \bibfield{author}{%
  \bibinfo {author} {\bibfnamefont{Harold~W.}\ \bibnamefont{Hatch}}, \bibinfo
  {author} {\bibfnamefont{Frank~H.}\ \bibnamefont{Stillinger}},\ and\ \bibinfo
  {author} {\bibfnamefont{Pablo~G.}\ \bibnamefont{Debenedetti}},\ }%
  \bibfield{title}{%
  \enquote{\bibinfo {title} {{Computational study of the stability of the
  miniprotein Trp-cage, the GB1 $\beta$-hairpin, and the AK16 peptide, under
  negative pressure}},}\ }%
  \bibfield{journal}{%
  \bibinfo {journal} {J. Phys. Chem. B}\ }%
  \textbf{\bibinfo {volume} {118}},\ \bibinfo {pages} {7761--7769} (\bibinfo
  {year} {2014})\BibitemShut{NoStop}%
\bibitem{mori-2014}%
  \BibitemOpen
  \bibfield{author}{%
  \bibinfo {author} {\bibfnamefont{Yoshiharu}\ \bibnamefont{Mori}}\ and\
  \bibinfo {author} {\bibfnamefont{Hisashi}\ \bibnamefont{Okumura}},\ }%
  \bibfield{title}{%
  \enquote{\bibinfo {title} {Molecular dynamics of the structural changes of
  helical peptides induced by pressure, dx.doi.org/10.1002/prot.24654},}\ }%
  \bibfield{journal}{%
  \bibinfo {journal} {Proteins}\ }%
  \textbf{\bibinfo {volume} {0}},\ \bibinfo {pages} {0--0} (\bibinfo {year}
  {2014})\BibitemShut{NoStop}%
\bibitem{best-2009-2}%
  \BibitemOpen
  \bibfield{author}{%
  \bibinfo {author} {\bibfnamefont{Robert~B.}\ \bibnamefont{Best}}\ and\
  \bibinfo {author} {\bibfnamefont{Gerhard}\ \bibnamefont{Hummer}},\ }%
  \bibfield{title}{%
  \enquote{\bibinfo {title} {Optimized molecular dynamics force fields applied
  to the helix-coil transition of polypeptides},}\ }%
  \bibfield{journal}{%
  \bibinfo {journal} {J. Phys. Chem. B}\ }%
  \textbf{\bibinfo {volume} {113}},\ \bibinfo {pages} {9004--9015} (\bibinfo
  {year} {2009})\BibitemShut{NoStop}%
\bibitem{jorgensen-1983}%
  \BibitemOpen
  \bibfield{author}{%
  \bibinfo {author} {\bibfnamefont{William~L.}\ \bibnamefont{Jorgensen}},
  \bibinfo {author} {\bibfnamefont{Jayaraman}\ \bibnamefont{Chandrasekhar}},\
  and\ \bibinfo {author} {\bibfnamefont{Jeffry~D.}\ \bibnamefont{Madura}},\ }%
  \bibfield{title}{%
  \enquote{\bibinfo {title} {Comparison of simple potential functions for
  simulating liquid water},}\ }%
  \bibfield{journal}{%
  \bibinfo {journal} {J. Chem. Phys.}\ }%
  \textbf{\bibinfo {volume} {79}},\ \bibinfo {pages} {926--935} (\bibinfo
  {month} {15 July}\ \bibinfo {year} {1983})\BibitemShut{NoStop}%
\bibitem{best-2010-3}%
  \BibitemOpen
  \bibfield{author}{%
  \bibinfo {author} {\bibfnamefont{Robert~B.}\ \bibnamefont{Best}}\ and\
  \bibinfo {author} {\bibfnamefont{Jeetain}\ \bibnamefont{Mittal}},\ }%
  \bibfield{title}{%
  \enquote{\bibinfo {title} {Protein simulations with an optimized water model:
  cooperative helix formation and temperature-induced unfolded state
  collapse},}\ }%
  \bibfield{journal}{%
  \bibinfo {journal} {J. Phys. Chem. B}\ }%
  \textbf{\bibinfo {volume} {114}},\ \bibinfo {pages} {14916--14923} (\bibinfo
  {year} {2010})\BibitemShut{NoStop}%
\bibitem{abascal-2005}%
  \BibitemOpen
  \bibfield{author}{%
  \bibinfo {author} {\bibfnamefont{J.~L.~F.}\ \bibnamefont{Abascal}}\ and\
  \bibinfo {author} {\bibfnamefont{C.}~\bibnamefont{Vega}},\ }%
  \bibfield{title}{%
  \enquote{\bibinfo {title} {A general purpose model for the condensed phases
  of water: Tip4p/2005},}\ }%
  \bibfield{journal}{%
  \bibinfo {journal} {J. Chem. Phys.}\ }%
  \textbf{\bibinfo {volume} {123}},\ \bibinfo {pages} {234505} (\bibinfo {year}
  {2005})\BibitemShut{NoStop}%
\bibitem{hess-2008}%
  \BibitemOpen
  \bibfield{author}{%
  \bibinfo {author} {\bibfnamefont{Berk}\ \bibnamefont{Hess}}, \bibinfo
  {author} {\bibfnamefont{Carsten}\ \bibnamefont{Kutzner}}, \bibinfo {author}
  {\bibfnamefont{David}\ \bibnamefont{van~der Spoel}},\ and\ \bibinfo {author}
  {\bibfnamefont{Erik}\ \bibnamefont{Lindahl}},\ }%
  \bibfield{title}{%
  \enquote{\bibinfo {title} {{GROMACS} 4: algorithms for highly efficient,
  load-balanced, and scalable molecular simulation},}\ }%
  \bibfield{journal}{%
  \bibinfo {journal} {J. Chem. Theory Comput.}\ }%
  \textbf{\bibinfo {volume} {4}},\ \bibinfo {pages} {435--447} (\bibinfo {year}
  {2008})\BibitemShut{NoStop}%
\bibitem{parrinello-1981}%
  \BibitemOpen
  \bibfield{author}{%
  \bibinfo {author} {\bibfnamefont{Michele}\ \bibnamefont{Parrinello}}\ and\
  \bibinfo {author} {\bibfnamefont{Aneesur}\ \bibnamefont{Rahman}},\ }%
  \bibfield{title}{%
  \enquote{\bibinfo {title} {Polymorphic transitions in single crystals: a new
  molecular dynamics method},}\ }%
  \bibfield{journal}{%
  \bibinfo {journal} {J. Appl. Phys.}\ }%
  \textbf{\bibinfo {volume} {52}},\ \bibinfo {pages} {7182--7190} (\bibinfo
  {year} {1981})\BibitemShut{NoStop}%
\bibitem{lifson-1961}%
  \BibitemOpen
  \bibfield{author}{%
  \bibinfo {author} {\bibfnamefont{Schneior}\ \bibnamefont{Lifson}}\ and\
  \bibinfo {author} {\bibfnamefont{A.}~\bibnamefont{Roig}},\ }%
  \bibfield{title}{%
  \enquote{\bibinfo {title} {On the theory of helix-coil transition in
  polypeptides},}\ }%
  \bibfield{journal}{%
  \bibinfo {journal} {J. Chem. Phys.}\ }%
  \textbf{\bibinfo {volume} {34}},\ \bibinfo {pages} {1963--1974} (\bibinfo
  {year} {1961})\BibitemShut{NoStop}%
\bibitem{zimm-1959}%
  \BibitemOpen
  \bibfield{author}{%
  \bibinfo {author} {\bibfnamefont{B.~H.}\ \bibnamefont{Zimm}}\ and\ \bibinfo
  {author} {\bibfnamefont{J.~K.}\ \bibnamefont{Bragg}},\ }%
  \bibfield{title}{%
  \enquote{\bibinfo {title} {Theory of phase transition between helix and
  random coil in polypeptide chains},}\ }%
  \bibfield{journal}{%
  \bibinfo {journal} {J. Chem. Phys.}\ }%
  \textbf{\bibinfo {volume} {11}},\ \bibinfo {pages} {526--535} (\bibinfo
  {year} {1959})\BibitemShut{NoStop}%
\bibitem{qian-1992}%
  \BibitemOpen
  \bibfield{author}{%
  \bibinfo {author} {\bibfnamefont{Hong}\ \bibnamefont{Qian}}\ and\ \bibinfo
  {author} {\bibfnamefont{John~A.}\ \bibnamefont{Schellman}},\ }%
  \bibfield{title}{%
  \enquote{\bibinfo {title} {Helix-coil theories: a comparative study for
  finite length polypeptides},}\ }%
  \bibfield{journal}{%
  \bibinfo {journal} {J. Phys. Chem.}\ }%
  \textbf{\bibinfo {volume} {96}},\ \bibinfo {pages} {3987--3994} (\bibinfo
  {year} {1992})\BibitemShut{NoStop}%
\bibitem{duan-2003}%
  \BibitemOpen
  \bibfield{author}{%
  \bibinfo {author} {\bibfnamefont{Yong}\ \bibnamefont{Duan}}, \bibinfo
  {author} {\bibfnamefont{Chun}\ \bibnamefont{Wu}}, \bibinfo {author}
  {\bibfnamefont{Shibasish}\ \bibnamefont{Chowdhury}}, \bibinfo {author}
  {\bibfnamefont{Matthew~C.}\ \bibnamefont{Lee}}, \bibinfo {author}
  {\bibfnamefont{Guoming}\ \bibnamefont{Xiong}}, \bibinfo {author}
  {\bibfnamefont{Wei}\ \bibnamefont{Zhang}}, \bibinfo {author}
  {\bibfnamefont{Rong}\ \bibnamefont{Yang}}, \bibinfo {author}
  {\bibfnamefont{Piotr}\ \bibnamefont{Cieplak}}, \bibinfo {author}
  {\bibfnamefont{Ray}\ \bibnamefont{Luo}}, \bibinfo {author}
  {\bibfnamefont{Taisung}\ \bibnamefont{Lee}}, \bibinfo {author}
  {\bibfnamefont{James}\ \bibnamefont{Caldwell}}, \bibinfo {author}
  {\bibfnamefont{Junmei}\ \bibnamefont{Wang}},\ and\ \bibinfo {author}
  {\bibfnamefont{Peter~A.}\ \bibnamefont{Kollman}},\ }%
  \bibfield{title}{%
  \enquote{\bibinfo {title} {A point-charge force field for molecular mechanics
  simulations of proteins based on condensed-phase quantum chemical
  calculations},}\ }%
  \bibfield{journal}{%
  \bibinfo {journal} {J. Comp. Chem.}\ }%
  \textbf{\bibinfo {volume} {24}},\ \bibinfo {pages} {1999--2012} (\bibinfo
  {year} {2003})\BibitemShut{NoStop}%
\bibitem{shalongo-1994}%
  \BibitemOpen
  \bibfield{author}{%
  \bibinfo {author} {\bibfnamefont{William}\ \bibnamefont{Shalongo}}, \bibinfo
  {author} {\bibfnamefont{Laxmichand}\ \bibnamefont{Dugad}},\ and\ \bibinfo
  {author} {\bibfnamefont{Earle}\ \bibnamefont{Stellwagen}},\ }%
  \bibfield{title}{%
  \enquote{\bibinfo {title} {{Distribution of helicity within the model peptide
  Acetyl(AAQAA)$_3$amide}},}\ }%
  \bibfield{journal}{%
  \bibinfo {journal} {J. Am. Chem. Soc.}\ }%
  \textbf{\bibinfo {volume} {116}},\ \bibinfo {pages} {8288--8293} (\bibinfo
  {year} {1994})\BibitemShut{NoStop}%
\bibitem{best-2012-2}%
  \BibitemOpen
  \bibfield{author}{%
  \bibinfo {author} {\bibfnamefont{Robert~B.}\ \bibnamefont{Best}}, \bibinfo
  {author} {\bibfnamefont{David}\ \bibnamefont{de~Sancho}},\ and\ \bibinfo
  {author} {\bibfnamefont{Jeetain}\ \bibnamefont{Mittal}},\ }%
  \bibfield{title}{%
  \enquote{\bibinfo {title} {Residue-specific {$\alpha$}-helix propensities
  from molecular simulation},}\ }%
  \bibfield{journal}{%
  \bibinfo {journal} {Biophys. J.}\ }%
  \textbf{\bibinfo {volume} {102}},\ \bibinfo {pages} {1897--1906} (\bibinfo
  {year} {2012})\BibitemShut{NoStop}%
\bibitem{kabsch-1983}%
  \BibitemOpen
  \bibfield{author}{%
  \bibinfo {author} {\bibfnamefont{Wolfgang}\ \bibnamefont{Kabsch}}\ and\
  \bibinfo {author} {\bibfnamefont{Christian}\ \bibnamefont{Sander}},\ }%
  \bibfield{title}{%
  \enquote{\bibinfo {title} {Dictionary of protein secondary structure: pattern
  recognition of hydrogen-bonded and geometrical features},}\ }%
  \bibfield{journal}{%
  \bibinfo {journal} {Biopolymers}\ }%
  \textbf{\bibinfo {volume} {22}},\ \bibinfo {pages} {2577--2637} (\bibinfo
  {year} {1983})\BibitemShut{NoStop}%
\bibitem{suppinfo}%
  \BibitemOpen
  \enquote{\bibinfo {title} {{See Supplementary Material Document No. XXX for a
  figure showing the quality of the thermodynamic model fit to the primary
  data. For information on Supplementary Material, see
  http://www.aip.org/pubservs/epaps.html}},}\ \BibitemShut{NoStop}%
\bibitem{paschek-2008}%
  \BibitemOpen
  \bibfield{author}{%
  \bibinfo {author} {\bibfnamefont{Dietmar}\ \bibnamefont{Paschek}}, \bibinfo
  {author} {\bibfnamefont{Sascha}\ \bibnamefont{Hempel}},\ and\ \bibinfo
  {author} {\bibfnamefont{Angel~E.}\ \bibnamefont{Garcia}},\ }%
  \bibfield{title}{%
  \enquote{\bibinfo {title} {Computing the stability diagram of the {Trp}-cage
  miniprotein},}\ }%
  \bibfield{journal}{%
  \bibinfo {journal} {Proc. Natl. Acad. Sci. U. S. A.}\ }%
  \textbf{\bibinfo {volume} {105}},\ \bibinfo {pages} {17754--17759} (\bibinfo
  {year} {2008})\BibitemShut{NoStop}%
\bibitem{roche-2013}%
  \BibitemOpen
  \bibfield{author}{%
  \bibinfo {author} {\bibfnamefont{Julien}\ \bibnamefont{Roche}}, \bibinfo
  {author} {\bibfnamefont{Mariana}\ \bibnamefont{Dellarole}}, \bibinfo {author}
  {\bibfnamefont{Jose~A.}\ \bibnamefont{Caro}}, \bibinfo {author}
  {\bibfnamefont{Douglas~R.}\ \bibnamefont{Norberto}}, \bibinfo {author}
  {\bibfnamefont{Angel~E.}\ \bibnamefont{Garcia}}, \bibinfo {author}
  {\bibfnamefont{Bertrand~E.}\ \bibnamefont{Garcia-Moreno}}, \bibinfo {author}
  {\bibfnamefont{Christian}\ \bibnamefont{Roumestand}},\ and\ \bibinfo {author}
  {\bibfnamefont{Catherine~A.}\ \bibnamefont{Royer}},\ }%
  \bibfield{title}{%
  \enquote{\bibinfo {title} {Effect of internal cavities on folding rates and
  routes revealed by real-time pressure-jump {NMR} spectroscopy},}\ }%
  \bibfield{journal}{%
  \bibinfo {journal} {J. Am. Chem. Soc.}\ }%
  \textbf{\bibinfo {volume} {135}},\ \bibinfo {pages} {14610--14618} (\bibinfo
  {year} {2013})\BibitemShut{NoStop}%
\bibitem{liu-2014}%
  \BibitemOpen
  \bibfield{author}{%
  \bibinfo {author} {\bibfnamefont{Yanxin}\ \bibnamefont{Liu}}, \bibinfo
  {author} {\bibfnamefont{Maxim~B.}\ \bibnamefont{Prigozhin}}, \bibinfo
  {author} {\bibfnamefont{Klaus}\ \bibnamefont{Schulten}},\ and\ \bibinfo
  {author} {\bibfnamefont{Martin}\ \bibnamefont{Gruebele}},\ }%
  \bibfield{title}{%
  \enquote{\bibinfo {title} {Observation of complete pressure-jump protein
  refolding in molecular dynamics simulation and experiment},}\ }%
  \bibfield{journal}{%
  \bibinfo {journal} {J. Am. Chem. Soc.}\ }%
  \textbf{\bibinfo {volume} {136}},\ \bibinfo {pages} {4265--4272} (\bibinfo
  {year} {2014})\BibitemShut{NoStop}%
\end{thebibliography}%


\begin{thebibliography}{1}%
\makeatletter
\providecommand \@ifxundefined [1]{%
 \@ifx{#1\undefined}
}%
\providecommand \@ifnum [1]{%
 \ifnum #1\expandafter \@firstoftwo
 \else \expandafter \@secondoftwo
 \fi
}%
\providecommand \@ifx [1]{%
 \ifx #1\expandafter \@firstoftwo
 \else \expandafter \@secondoftwo
 \fi
}%
\providecommand \natexlab [1]{#1}%
\providecommand \enquote  [1]{``#1''}%
\providecommand \bibnamefont  [1]{#1}%
\providecommand \bibfnamefont [1]{#1}%
\providecommand \citenamefont [1]{#1}%
\providecommand \href@noop [0]{\@secondoftwo}%
\providecommand \href [0]{\begingroup \@sanitize@url \@href}%
\providecommand \@href[1]{\@@startlink{#1}\@@href}%
\providecommand \@@href[1]{\endgroup#1\@@endlink}%
\providecommand \@sanitize@url [0]{\catcode `\\12\catcode `\$12\catcode
  `\&12\catcode `\#12\catcode `\^12\catcode `\_12\catcode `\%12\relax}%
\providecommand \@@startlink[1]{}%
\providecommand \@@endlink[0]{}%
\providecommand \url  [0]{\begingroup\@sanitize@url \@url }%
\providecommand \@url [1]{\endgroup\@href {#1}{\urlprefix }}%
\providecommand \urlprefix  [0]{URL }%
\providecommand \Eprint [0]{\href }%
\providecommand \doibase [0]{http://dx.doi.org/}%
\providecommand \selectlanguage [0]{\@gobble}%
\providecommand \bibinfo  [0]{\@secondoftwo}%
\providecommand \bibfield  [0]{\@secondoftwo}%
\providecommand \translation [1]{[#1]}%
\providecommand \BibitemOpen [0]{}%
\providecommand \bibitemStop [0]{}%
\providecommand \bibitemNoStop [0]{.\EOS\space}%
\providecommand \EOS [0]{\spacefactor3000\relax}%
\providecommand \BibitemShut  [1]{\csname bibitem#1\endcsname}%
\let\auto@bib@innerbib\@empty
\bibitem [{\citenamefont {Imamura}\ and\ \citenamefont
  {Kato}(2008)}]{imamura-2008}%
  \BibitemOpen
  \bibfield  {author} {\bibinfo {author} {\bibfnamefont {H.}~\bibnamefont
  {Imamura}}\ and\ \bibinfo {author} {\bibfnamefont {M.}~\bibnamefont {Kato}},\
  }\bibfield  {title} {\enquote {\bibinfo {title} {{Effect of pressure on
  helix-coil transition of an alanine-based peptide: An FTIR study}},}\
  }\href@noop {} {\bibfield  {journal} {\bibinfo  {journal} {Proteins}\
  }\textbf {\bibinfo {volume} {75}},\ \bibinfo {pages} {911--918} (\bibinfo
  {year} {2008})}\BibitemShut {NoStop}%
\end{thebibliography}%
\end{document}


\title{Supporting Information For: 
Role of solvation in pressure-induced helix stabilization}

\author{Robert B. Best}
\email{robertbe@helix.nih.gov}
\affiliation{Laboratory of Chemical Physics,
National Institute of Diabetes and Digestive
and Kidney Diseases, National Institutes of Health, Bethesda, MD 20892-0520, U.S.A.}
\author{Cayla Miller}
\affiliation{Department of Chemical and Biomolecular Engineering, Lehigh University, Bethlehem, PA 18015, U.S.A.}
\author{Jeetain Mittal}
\email{jeetain@lehigh.edu}
\affiliation{Department of Chemical and Biomolecular Engineering, Lehigh University, Bethlehem, PA 18015, U.S.A.}


\maketitle

\begin{figure}[h]
\centering
\begin{tabular}{c}
\includegraphics[width=7cm]{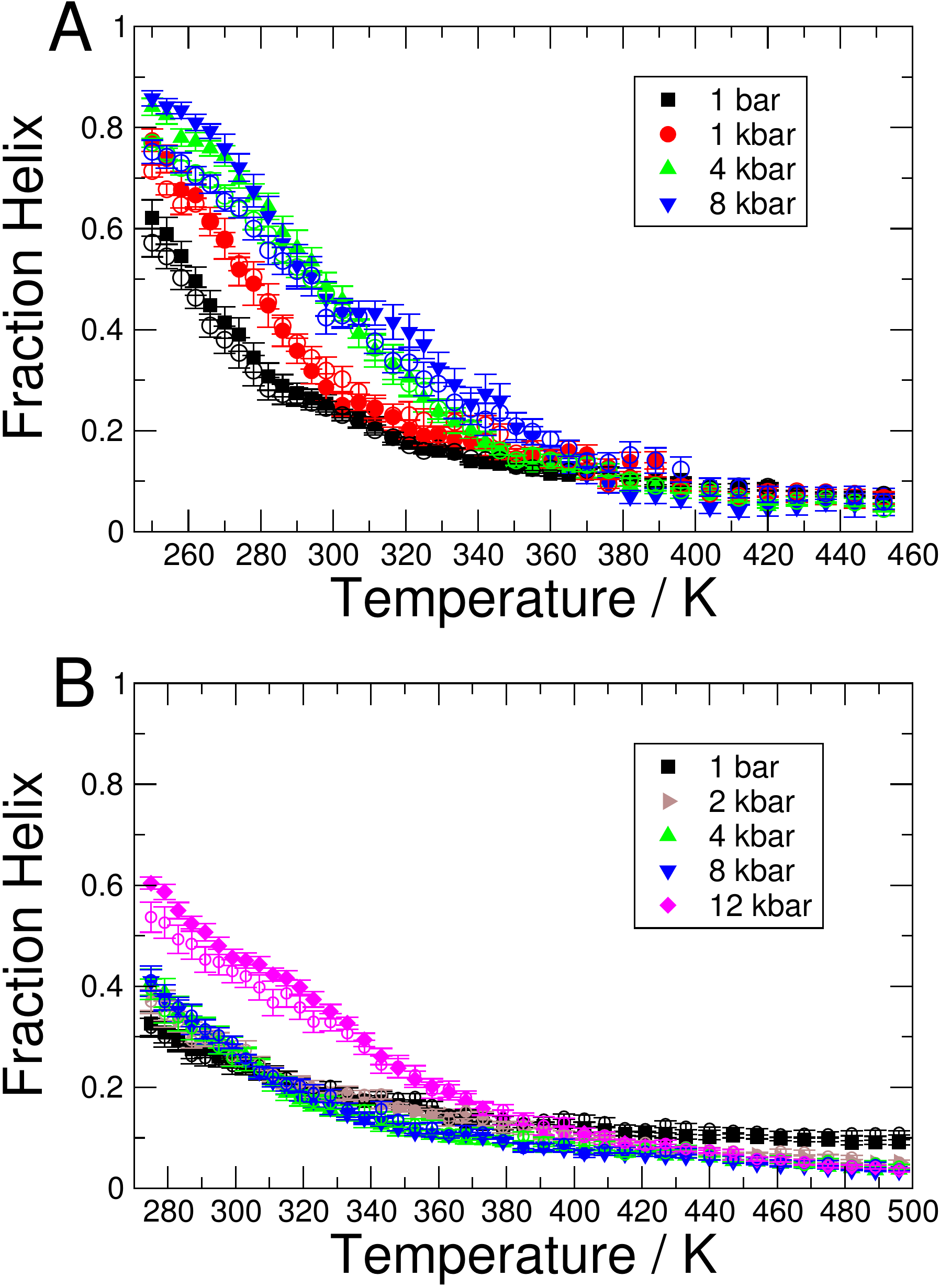}
\end{tabular}
\caption{Alternative definition of fraction helix. The 
temperature-dependent helix formation from REMD at 
constant pressure is shown for (A) Amber ff03w and
(B) Amber ff03*. Solid symbols are for helix fraction 
determined from torsion angles (see main text) and 
open symbols with matching colors are for helix fraction
determined from DSSP\cite{kabsch-1983} using 
$f_\mathrm{helix}=n_\alpha/n_\mathrm{res}$, where
$n_\alpha$ is the number of helical residues from DSSP and
$n_\mathrm{res}\equiv15$ is the number of residues
in the peptide. All other details are as in Fig. 1 in the
main text.}
\label{si-fit-fig}
\end{figure}

\begin{figure}[h]
\centering
\begin{tabular}{c}
\includegraphics[width=7cm]{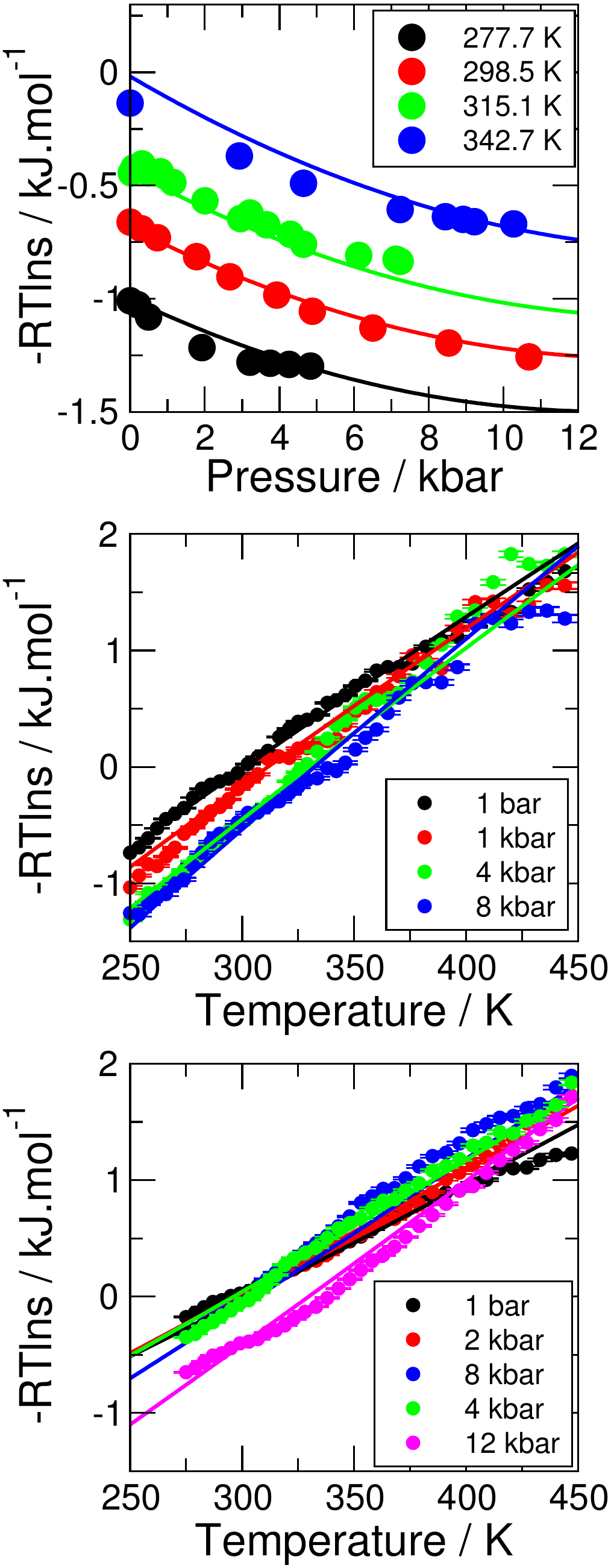}
\end{tabular}
\caption{Fit of thermodynamic model to raw data.
(Top) Experimental data from Imamura and Kato\cite{imamura-2008};
(Center) Simulation data for Amber ff03w; 
(Lower) Simulation data for Amber ff03*. Symbols are 
data, lines with corresponding colors are fits to  the
thermodynamic model. }
\label{si-fit-fig}
\end{figure}

\clearpage

\bibliography{references}